\documentclass[12pt, preprint]{aastex}
\usepackage{geometry,amsmath}
\usepackage{natbib}
\newcommand\he{10830~\AA{}}

\shortauthors{Zeng et al.}
\shorttitle{}
\begin{document}

\title{\uppercase{A Flare Observed in Coronal, Transition Region and Helium~I \he{} emissions}}
\author{Zhicheng Zeng$^{1, 2}$, Jiong Qiu$^{3}$, Wenda Cao$^{1, 2}$, and Philip G. Judge$^{4}$}
\affil{1.Center for Solar-Terrestrial Research, New Jersey Institute of Technology, 323 Martin Luther King Blvd., Newark, NJ 07102, USA \\
2. Big Bear Solar Observatory, 40386 North Shore Lane, Big Bear City, CA 92314, USA\\
3. Department of Physics, Montana State University, Bozeman, MT 59717-3840, USA\\
4. High Altitude Observatory, National Center for Atmospheric Research, Boulder CO 80307-3000, USA }

\begin{abstract}
  On June 17, 2012, we observed the evolution of 
 a C-class flare associated with the eruption of a filament near 
  a large sunspot in the active region NOAA 11504.  We obtained high spatial resolution
  filtergrams using the 1.6 m New Solar Telescope at the Big Bear Solar
  Observatory in broad-band TiO at 706~nm (bandpass: 10~\AA)
  and \ion{He}{1} \he{} narrow-band (bandpass: 0.5~\AA, centered 0.25~\AA{} to the
  blue). 
   We analyze the spatio-temporal behavior of the \ion{He}{1} \he{} data, 
which were obtained over
  a 90\arcsec$\times$90$\arcsec$  field of view with a cadence of 10 sec.
  We also analyze simultaneous data from the Atmospheric Imaging Assembly 
  and Extreme Ultraviolet Variability Experiment instruments on board the Solar Dynamics Observatory spacecraft,
   and data from Reuven Ramaty High Energy Solar Spectroscopic Imager and GOES
  spacecrafts. Non-thermal effects are ignored in this analysis. 
  Several quantitative aspects of the data, as well as 
 models derived using the ``0D'' Enthalpy-Based Thermal Evolution of Loops model \citep[EBTEL:][]{Klimchuk08} code, 
 indicate that the triplet states of the \he{}
    multiplet are populated by photoionization of chromospheric
  plasma followed by radiative recombination.  Surprisingly, the
  \ion{He}{2} 304~\AA{} line is reasonably well matched by standard
  emission measure calculations, along with the \ion{C}{4} emission
  which dominates the AIA 1600~\AA{} channel during flares.  This
  work lends support to some of our previous work combining X-ray, EUV and UV
  data of flares to build models of energy transport from corona to chromosphere.
\end{abstract}


\section{INTRODUCTION}

The spectra of helium atoms and ions in the Sun are not yet
understood.  The EUV resonance lines at \ion{He}{1} 584~\AA{} and \ion{He}{2} 304~\AA{}
respectively are anomalously bright, under quiescent conditions, by
factors of several when compared with many other lines \citep{Jordan75, Jordan80, Zirin75, Pietarila04, Judge04}.
Yet, these lines are
some of the strongest EUV features in the solar spectrum, and as such they
control to a significant degree the state of the earth's thermosphere
and ionosphere.  Until a clear understanding of the formation of these
lines is available, attempts to model the EUV irradiance using models
based upon the standard assumption of ionization equilibrium are
doomed to fail.  In the present study we analyze the evolution of
flare emissions in EUV and UV radiation and emission from neutral and
ionized helium, to probe the mechanisms leading to the strong helium
emission.   A broader goal of the present paper is to understand the evolution of the
emitting plasmas during flares.  We use the narrow-band tunable filtergraph for
the New Solar Telescope at the
Big Bear Solar Observatory \citep[NST/BBSO;][]{Goode12, CaoWD10b} to capture flare emission in the \he{} line.  By
combining these data with those of the adjacent chromosphere and
corona seen at UV and EUV wavelengths with the Atmosphere Imaging Assembly \citep[AIA;][]{Lemen12} and 
Extreme Ultraviolet Variability Experiment \citep[EVE;][]{Woods12} instruments on the {\it Solar Dynamic Observatory} \citep[{\it SDO};][]{Pesnell12}, we use the observed
behavior to constrain the mechanisms by which helium emission can
occur during flares.

Ground-based observations of helium lines that follow the 
evolution of flares are not common, the emission being confined to narrow kernels that
are ill-suited to observation using slit spectrographs.  Some
important work by \citet{Tandberg67} found that in the weaker
flares \ion{He}{1} \he{} shows absorption and only class 2B and
larger flares show \ion{He}{1} \he{} emission. Later, \ion{He}{1}
\he{} emissions were observed in a C9.7 flare with
spectro-polarimetry \citep{Penn95}. Using an infrared (IR) spectrograph with a spatial resolution of $1\farcs34$ 
and a temporal cadence of 2.8 s, \citet{LiH07} found that only when the GOES X-ray flux (which is 
integrated over the solar disk) reaches a threshold (about C6 class in their study) could they
detect emission exceeding the continuum, which spatially corresponded
to a bright H$\alpha$ kernel.

\citet{Kleint12} has analyzed
data of photospheric \ion{Fe}{1} 6302~\AA{} and 
chromospheric \ion{Ca}{2} 8542~\AA{} lines in a C3 class flare. She  finds 
that emission occurs only above the photosphere, within chromospheric
plasma.   It
should be remembered that the chromosphere spans some 9 pressure scale
heights between the quiescent (pre-flare) photosphere and overlying
corona.   These facts are suggestive that the \ion{He}{1} \he{} emission
should also arise from the chromospheric layers inside the flare footpoints.

Any process that can excite helium leading to line 
radiation requires a change in principle quantum number
$n$ from the $n=1$ ground levels.  Thus helium excitation is unusually
sensitive to {\em high
energy tails in distribution functions of the radiation field and/or
plasma electrons and ions}.  For example, the resonance lines ($2s-2p$) of
\ion{C}{4} have a 5x lower threshold than the resonance line of
\ion{He}{2} ($1s-2p$) which form at similar temperatures, under ionization
equilibrium conditions.  One  proposal uses ``high energy
photons'' (shortward of 228 and 504~\AA{}) to ionize ionized/neutral helium from which
recombination populates the $n=2,3$ levels of helium \citep{Zirin75}. 
By solving the standard non-LTE radiation transfer problem in 1D semi-empirical models of the solar atmosphere, 
it is found that the properties of the \ion{He}{1} \he{} multiplet depend mainly 
on the density and the thickness of the chromosphere, 
as well as on the incoming EUV coronal back-radiation. \citep{Pozhalova88, Avrett94, Centeno08}. 

Another appeals to the diffusion of helium atoms and ions up
steep temperature gradients where the locally higher temperature can
excite the $n=2$ levels faster than occurs for transitions such as the
\ion{C}{4} resonance lines \citep{Jordan75}.  Yet another uses unresolved
plasma motion to achieve the same result \citep{Andretta00}.   
Flare spectra of EUV helium emission lines,
observed with the SO82A (spectroheliograph, or ``overlap-ogram'')
instrument, were analyzed in
terms of a ``burst'' model by \citet{Laming92}. \cite{Laming92}
 assumed that,  during flares, ``ionizing plasma'' conditions 
exist in the Sun's transition region, permitting the excitation of
$n=2,3$ levels before ionization occurs.   \citet{Pietarila04}
showed that all of these mechanisms can be unified
into a common picture of excitation, by considering the conditions
experienced by the motions of 
individual atoms and ions of helium (Lagrangian
picture).

No matter the precise mechanism at play, helium may be expected to
shed light on processes requiring much higher energies ($> 10$ eV) than
are present throughout the bulk of the chromosphere ($\sim 0.6$ eV).  Near
simultaneous observations of various photospheric and chromospheric
lines show that the chromosphere is the location of the footpoint emission
from at least moderate (C-class) flares \citep{Kleint12}.

Furthermore, the importance of He~I 10830 triplet is being realized from 
monitoring dynamical activity in the chromosphere and measuring the chromospheric magnetic fields.
 \citet{Trujillo07} have modelled the scattering polarization by interpreting the observed polarization in terms of the Zeeman
 and Hanle effects.
 
We proceed by combining high angular resolution observations of the \ion{He}{1}
\he{} multiplet with lower resolution measurements of \ion{He}{2} 304~\AA{}
obtained with SDO.  Other observables from SDO, as well as X-ray emissions
from the {\it Reuven Ramaty High Energy Solar Spectroscopic Imager} \citep[{\it RHESSI};][]{LinRP02} 
and GOES will be used to understand the changing conditions in the emitting
chromospheric and coronal plasma.

\section{OBSERVATIONS}

With the advent of the 1.6 meter NST at the BBSO, ground based IR observations with high resolution
($\dfrac{\lambda}{D}\sim0\farcs14$ in \he{} are now regularly acquired. The off-axis design of NST
reduces stray light since there is no central obscuration, and the
site's good seeing conditions combined with the high order adaptive optics (AO)
have enabled us to obtain observations with a spatial resolution close
to the diffraction limit of the telescope. On June 17, 2012, we observed 
a small filament eruption and the associated flaring activity.    
One footpoint of the filament was 
embedded in a small pore near a large sunspot in the active region
NOAA 11504.  High spatial resolution filtergrams were obtained in a
broad band (bandpass: 10~\AA) containing well-known TiO  lines, and in 
a narrow band 
(bandpass: 0.5~\AA) in the blue wing of the 
\ion{He}{1} \he{} multiplet.  The latter images 
enabled us to establish the photospheric underpinning of the filament in the chromosphere and corona. 

\subsection{Infrared Data}

Our TiO images have a cadence of 30 seconds,
with a field-of-view (FOV) of $70\arcsec \times 70\arcsec$. The image scale is 0\farcs0375/pixel. TiO molecular bands are unusually 
sensitive to
the photospheric temperature, they exhibit enhanced intensity contrast in the
photosphere due to the stronger absorption of these bands in 
cool intergranular lanes. The \he{} Lyot filter was made by the Nanjing Institute for 
Astronomical and Optical Technology. This narrow-band filter was tuned
to -0.25~\AA{} relative to the two blended strongest components of
the multiplet (at 10830.3~\AA), making the
filtergram more sensitive to upward moving features. A high
sensitivity HgCdTe CMOS IR focal plane array camera \citep{CaoWD10a} 
was employed to acquire the \he{} data with a cadence
of 10 seconds.  The pixel size of the filtergrams is 
0\farcs0875, and the FOV is 90$''$ by 90$''$.  With the aid of high
order AO \citep{CaoWD10b} and speckle reconstruction post-processing
\citep[KISIP speckle reconstruction code,][]{Woger07},
images with diffraction limited resolution at different bands were
achieved. All NST data used in this paper are speckle reconstructed. 
More technical detail in data acquisition and reduction can be found in \citet{Cao10c}.

With the NST we observed the active region continuously for a few hours in good
seeing conditions, providing a unique opportunity to follow
the entire evolution of the flare with high spatial and
temporal resolutions. A movie (NST\_flare\_20120617.mpg) is available in on-line journal.  Observations started before 17:00 UT, well
before the flaring activity.  Figure~1 is a snapshot of the
\he{} filtergram during the maximum of the flare, with the
bandpass shown in the lower right panel. Since the 
bandpass is tuned to the blue wing of \he{},  the
filtergram captures some underlying continuum radiation, with photospheric
features including sunspots, pores, and granules clearly visible in the
image.  The filament appears as a dark feature lying between two large
sunspots. During the flare, impulsive brightenings in this bandpass
are observed at a few locations, marked as patches 0 to 8 in the
figure; these patches will be described as P0 - P8 in the following
analysis.

\subsection{Supporting Data from SDO and RHESSI}

We also take advantage of the SDO's unique continuous
coverage of the Sun.  We analyze UV and EUV
images observed by AIA, and full disk continuum images and
line-of-sight (LOS) magnetograms from the Helioseismic and Magnetic
Imager \citep[HMI;][]{Schou12}. AIA takes full\--disk images
in 7 EUV bands with a cadence of 12~s and spatial scale of
0\farcs6. These EUV telescopes observe emission from coronal plasmas
at temperatures from 1 to 10 MK. Apart from these EUV channels, AIA
also takes full\--disk images at the UV 1600~\AA{} broadband at 24~s cadence. 
During flares, the \ion{C}{4} doublet is significantly enhanced to dominate
emission in this band \citep{QiuJ12}, also confirmed in the present
analysis. 
The \ion{C}{4} line emission is produced by
plasmas near 100,000 K, typical of the mid transition region. 

Spatial co-alignment among these data is done by first co-aligning the
HMI continuum images with the NST TiO data. Using the granular
patterns and sunspots, 
it is straightforward to
precisely align the HMI/SDO continuum images with the NST TiO
images. Then we co-align the TiO data with the \he{} filtergrams
using the sunspots and bright granules. AIA and HMI
observations are co-aligned with each other using satellite pointing
information, and the HMI continuum images serve as the
intermediary for co-alignment between the AIA images and the NST
images.  The accuracy of the co-alignment is better than $0\farcs5$.

Finally, X-ray emission of the flare are also observed by GOES and RHESSI.
Using the standard RHESSI software package, we obtain RHESSI X-ray light curves and images with pixel size of 2.26$\arcsec$ by 2.26$\arcsec$ (subcollimators 1, 3, 4, 5, 6, 8 are used).
The RHESSI data are co-aligned with the SDO data according to the coordinates provided in the FITS header.
 Light curves and images of the flare
in the X-ray, EUV, UV, and IR wavelengths are illustrated in Figure~2
and Figure~3 respectively, derived as discussed below.

\subsection{Overview of the Flare}

In the top panel of Figure~2, X-ray light curves in several RHESSI
channels and in GOES 1-8~\AA{} channel are plotted\footnote{Note that
  10 keV corresponds to a photon wavelength of 1~\AA.}. 
  Colored curves in panel (b) of Figure~2 show normalized light curves in several wavelengths:
  the IR \he{} from NST, EUV bands at 94~\AA\ and 171~\AA{}
  from AIA/SDO, and UV band at 1600~\AA{} also from AIA/SDO.  In panel~(b), the 1600~\AA\ light curve (green) is created from total counts of selected footpoint pixels
   that exhibit strong emission, or more specifically those pixels with a count rate greater than five times the median count rate ($I_q$ = 71 $DN s^{-1}$) of the quiescent region. Since the AIA data and NST data have been well co-aligned, the light curve of \ion{He}{1} \he{} (black) is made from the counterpart of the footpoint pixels in 1600~\AA. The light curves in EUV 94~\AA\ (blue) and 171~\AA{} (red) are made from pixels of the entire FOV shown in Figure~3.
  Each curve is subtracted by its minimum values and then divided
  by the residual maximum. 
  As seen in Figure~2, flaring occurs in two phases, with the
  SXR emission in GOES 1-8~\AA{} peaking at 17:28 UT and 17:39 UT,
  respectively. Each phase is usually characterized by a rapid rise of the
  emission, followed by a more gradual
  decay. Remarkably, we find that emission in the IR and UV bands
  peaks first, along with a peak in the 171~\AA{} channel, 
  immediately followed by SXR emissions by relatively
  high temperature plasmas, and then by lower temperature EUV emissions
  in 94~\AA{} (6~MK) and 171~\AA{} (1~MK). For the EUV 171~\AA\
  light curve, the first two peaks coincide with the UV and IR peaks 
  while the following ones correspond to the cooling of the post-flare coronal loops. 
  
Figure~3 shows the evolving morphology of flaring plasma observed in
several wavelengths. 
Before the flare onset, twisting of the dark filament can be seen in the
\he{} filtergrams (see panel a1 of Figure~3 and the on-line movie (NST\_flare\_20120617.mpg)).  Before the flare is seen in X-ray,
 the filament is dark in \he{} filtergram as shown in
panel~a1 and there are no obvious signals in EUV images. However, in
UV 1600~\AA{} images (b1), there are brightenings right in the middle of the
filament, indicating the ongoing activity at upper chromosphere prior to the filament
eruption and flare. These small events account for a minor emission
peak at 17:22 UT before the flare as shown in Figure~2. 
Next, the filament brightening and eruption (second column in Figure~3) are
 coincident in time
with the first major abrupt rise in the light curves.  Ten minutes later, a
second abrupt brightening phase (third column in Figure~3) was observed as the emission of the
footpoint dramatically increases. The intensity of \he{} is
significantly enhanced by more than 50\% of pre-flare state.

In the top panels of Figure~3, contours of the X-ray images by RHESSI
are overlaid on the \he{} images. Strong soft X-ray (SXR)
emission at photon energies from 6-15~keV (green contours) is observed during
both phases of the flare, from an extended coronal emission source
lying above the foot-point patches significantly larger than the
SXR RHESSI angular resolution ($\sim 2-3\arcsec$), also shown in Figure~3(a2-a3). There is also a small
amount of X-ray emission above 25~keV, most likely the hard X-ray (HXR) emission (red contours), which is present for
only a short time. So these weak signals above 25~keV suggest that the non-thermal effects are not significant in this event  \citep[section 2.4]{Fletcher11}. The SXR emissions below 25~keV and in GOES 1-8~\AA{} channel are usually
produced by flare plasmas heated to over 10~MK \citep{Hannah08}.

During these two major phases of
the flare, emissions in IR, UV, and EUV bands occur in the
localized patches, P0-P8, which are connected by coronal loops
brightened later on in 131~\AA{} (e4) and 171~\AA{} (d4). 
Therefore, these IR and UV bright kernels most likely correspond to footpoints of
flare loops.  The counterpart in the
corona is shown in both 171~\AA{} and 131~\AA{} images (d1-e4),
reflecting the loop structures with million degree plasma. EUV flux observed by AIA are
saturated in 171~\AA{} and 131~\AA, especially during the first
impulsive phase. In the first impulsive phase, the saturations are in
the middle of the filament which is shown in the column a2-e2 of Figure~3,
and they coincide with both the RHESSI SXR contour and chromospheric
brightenings in 1600~\AA{} and 10830~\AA. Then during the second peak,
the footpoints become bright. These observations strongly suggest that the filament eruption
starts from chromosphere (early activity in AIA 1600~\AA\ inside filament) and when magnetic reconnection happens,
the released energy heats the plasma to the coronal temperature (SXR and EUV brightening in the first impulsive phase).
 Later, magnetic reconnection takes place in the disturbed corona, giving rise to more energy release
 and formation of new coronal loops as well as brightened foot-points; this generates the second phase of the
 flare. (The whole process is also seen in the on-line movie.).

Figure~4 shows the TiO images during the flare as well as one HMI
photospheric magnetogram, showing only LOS components of the
magnetic field.  Blue contours superimposed in these images are IR
foot-point patches brightened during the flare. The TiO images exhibit
lots of small-scale bright points, but barely any enhanced flare
emission at the locations of the bright IR patches.  
Therefore, we can assert that {\em the foot-point
  brightening observed in the He~\he{} band is not produced in
  the photosphere}. The panel (c)
  in Figure~4 shows one (typical) HMI LOS magnetogram. The blue and red
  contours are footpoint patches of the \he{} filtergram and
  coronal loop of the AIA 131~\AA{} image, respectively. Further, the IR footpoint patches are located
 in penumbral magnetic fields with different
polarity and most of them are seen to be connected by coronal loops. Patches 
P0, P1, P2, and P5 are located in magnetic fields of negative
polarity, while P3, P4, P6, P7, and P8 are located in positive
polarities. The loops in the  AIA 131~\AA{}
seem to connect  P0-P2 to  P3-P4, and P5 appears to be
connected with P6.  These loops are significantly inclined relative to the
magnetic polarity inversion line.

\subsection{Light Curves of Flare Foot-point Emission}

In Figure~5, we plot light curves of P0, P4,
and P6 in 10830~\AA, AIA 1600~\AA{} and 304~\AA, these being
typical examples. The GOES 1-8~\AA{} light curve is not plotted since it is not spatially resolved.
The patches are rectangular areas containing a combined 906 AIA pixels
(each with scale of 0\farcs6) that are centered over the enhanced 
\he{} regions 
brightened during the
flare (see Fig.~1). In Figure~5 we have subtracted
minimum values for each curve and then divided the curves by their residual maximum.   

In this paper, the IR filtergram contains both line wing and part of the line center of the \ion{He}{1} \he{} and the line profile before the flare is in absorption. 
Approximations at the footpoint are made as follows: 
\begin{enumerate}
\item{} Before the flare, the clear view of the penumbra indicates the absorption is optically thin and only a small fraction of continuum flux is absorbed.
Hence the observed flux before flare may be approximated as continuum.
\item{} The flare in this paper is not a white-light flare so the quiescent (pre-flare) continuum is the same as continuum during flare, which means that the emission enhancement is due to \ion{He}{1} \he{}; 
\item{} Therefore we assume that the flare enhancement with respect to pre-flare flux is approximately the same as enhancement over the quiescent continuum. 
\end{enumerate}
Figure~6 shows the histogram of the peak intensity, normalized
to the pre-flare emission, of these 906 pixels in different
wavelengths. The UV 1600~\AA{} and EUV 304~\AA{} flare
emissions have increased by up to two orders of magnitude, and the IR
intensity has grown by a factor of 1.2 to 2.5 over the pre-flare
emission. It is noteworthy that, previous flare observations in the \ion{He}{1}
10830~\AA{} bandpass have shown IR darkening \citep{Harvey84} and 
only when the GOES X-ray flux reaches a threshold could they detect emission \citep{Tandberg67, LiH07}.
However, in this flare, almost all flaring pixels exhibit IR emission 
even during the first phase when GOES 1-8~\AA\ emission is only at C1 level. 

As shown in Figure~5, the emissions from P0, P4, P6 in 10830~\AA, 1600~\AA{} and
304~\AA{} all rise rapidly on the same timescale, and peak at the same
time; they then decay more gradually but with quite different
timescales: the 304~\AA{} emission decays slightly slower than the
1600~\AA{} emission, and the \he{} IR intensity decays most slowly. Similar 
behavior is seen in individual pixels but with lower signal to noise
ratios. The statistics of rise and decay
timescales of all pixels are plotted in Figure~7.  
 For each pixel, the maximum of light curve is recorded. Then the light curve 
from the onset of the flare to the time of maximum is fitted using a half Gaussian function. 
The rising timescale is calculated as half of the full width at half maximum of the fitted Gaussian function. 
Then the curve from maximum is fitted to an e-slope.
The decay timescale is calculated from the time of its maximum to the time when it decays to $\dfrac{1}{e}$ of its maximum.

The flux of most pixels rise impulsively within one minute. 
The UV 1600~\AA{} flux typically decays within a few
minutes while the decay timescale of the 304~\AA{} flux is around 10
minutes. The flux in \he{} decays on longer timescales
of a few tens of minutes. 

\section{ANALYSIS}

\subsection{Summary of Critical Observations}

There are several critical aspects to IR \he{} narrow-band
and other data (refer to Figures 1-7) which we highlight and 
will draw upon below:

\begin{enumerate}

\item{}  The \he{} emission is confined to patches
  that are mophologically a mixture of bright fibril-like structures
  (P1, P5, P7), and bright amorphous patches (P0, P2, P3, P6, P8) (Figure~1). 

\item{} These patches lie mostly over penumbral regions, on large
  scales they are associated with bright 
 UV/EUV emission (Figure~4).

\item{} The \he{} emission often has sharp edges (e.g. P0), close to the
  resolution limit of the observations ($0\farcs14$) (Figure~1).  

\item{} The SXR emission observed below 15~keV by RHESSI is from a much extended source: in contrast to the sharp edges in \he{} (Figure~3). 

\item{} The simultaneous steep rise at 304, 1600 and \he{}
  footpoint emissions suggests that the IR and UV/EUV
  emissions are related during this phase (Figure~5).

\item{}  The \he{} emission decays an order of magnitude more slowly
than other UV/EUV emission, but has some similarity to the RHESSI and
GOES decay curves (Figure~2). 

\end{enumerate}

\subsection{Physical picture}

In previous work \citet{QiuJ12} have found a simple physical picture which can
account for the UV and EUV data of flares that are qualitatively
similar to the new observations presented here.  Some process to be
identified transfers energy rapidly out of the corona. When this
energy is directed towards the lower solar atmosphere then
this atmosphere responds by trying to deal with the excess energy
through ionization, radiation losses, fluid motions and perhaps other
modes (MHD waves, particle acceleration).  

Evolution of plasma at the feet of flare loops begins with a highly
dynamic response to impulsive heating.  In work by \citet{QiuJ12},
the ``impulsive rise''\footnote{Not to be confused with the
  ``impulsive'' phase of many flares seen at radio and gamma ray wavelengths.}  of
the UV 1600~\AA{} light curves at the flare foot-points is used to
estimate empirical heating rates through a ``0D'' model.  The
method, using only two free parameters, has been applied to analyze
and model heating of thousands of flare loops in a few flares with
agreeable comparison between synthetic and observed X-ray and EUV
spectra and light curves \citep{QiuJ12, LiuWJ13, LiY14}.  
Using UV signatures to build heating rates, these studies
not only resolve heating in individual loops but are not confined to
flares that have significant thick-target HXR emissions.  In
the present paper, we make two further simplified assumptions based
upon
previous work \citep{QiuJ12} and upon the weakness of HXR emissions: (1)
the 1600~\AA{} emission during the flare is dominated by \ion{C}{4}
emission and not the underlying continuum; (2) there are no
significant non-thermal particles impacting the solar chromosphere
from above.

The thermal conduction from the site of energy deposition 
will generate a downward propagating shock front, which most likely produces 
the impulsive spike in the UV and EUV light curves \citep{Fisher89}.
On the other hand, the cooling of the overlying corona governs the gradual decay. 
Such two phase evolution in the UV emissions has been reported by 
\citet{Hawley03} in stellar flare observations, 
and they found that during the cooling phase, a few lines including \ion{C}{4} can be
used as a transition region pressure gauge monitoring the coronal plasma evolution 
in overlying flare loops. The differential emission measure (DEM) throughout the
transition region is proportional to the equilibrium coronal pressure since
 the entire loop is in approximate hydrostatic balance in the cooling phase 
 \citep{Fisher87, Griffiths98, Hawley92}. In this ``pressure-gauge'' approximation, the decay phase of UV/EUV emission
in solar flares \citep{LiuWJ13, QiuJ13} also compares favorably with the observations.  Here, we adopt
this model to compute decay-phase flare foot-point emission in EUV
\ion{He}{2} 304~\AA, and UV \ion{C}{4} bands.  The IR \ion{He}{1} \he{} 
emission is then studied in the context of this model.  Below, we
model plasma evolution in hundreds of flare loops observed in this
flare to find the coronal/transition region structure overlying the
IR patches. We then estimate \ion{He}{1} \he{} enhancement due
to different physical processes in these patches to compare with
observations.

\subsection{Calculations with EBTEL}

For loops during the decay phase in approximate hydrostatic
balance, the DEM throughout the
transition region is proportional to the equilibrium coronal pressure 
and the optically thin radiative losses are balanced by the downward conductive heat from the cooling
coronal loops. With the plasma flow neglected, the DEM along the leg of
the flux tube is computed analytically \citep{Fisher87, Griffiths98, Hawley92}
as \newline

\begin{center} $\xi_{se}(T)
=\overline{P}\sqrt{\dfrac{\kappa_{0}}{8k^{2}_{B}}}T^{\frac{1}{2}}
Q^{-\frac{1}{2}}(T)$
\end{center} where
\begin{center}
$Q(T)=\int\limits_{T_0}^{T}T'^{\frac{1}{2}}\Lambda(T')dT'$
\end{center} and $\Lambda(T)$ is the optically thin radiative loss
function. $\kappa_{0}$ is the thermal conductivity coefficient and $k_{B}$
is Boltzmann constant. Expressing the temperature-dependent scaling constant as
$g_{se}(T)$, the transition region DEM can be computed as
$\xi_{se}(T)=g_{se}(T)\overline{P}$, which is directly proportional to
the mean pressure $\overline{P}$ calculated using the zero-dimensional loop heating model, the enthalpy-based
thermal evolution of loops model \citep[EBTEL;][]{Klimchuk08, Cargill12a, Cargill12b}.
 \citet{QiuJ13} and \citet{LiuWJ13} have already
confirmed that this pressure-gauge approximation can re-produce
observed gradual decay reasonably well.

We have used the spatially
resolved AIA 1600~\AA{} light curves to determine the empirical
heating rates of the observed flare loops. With these heating
rates, a zero-dimensional EBTEL model is used to compute mean temperature and density
of plasmas in these flare loops. EBTEL
solves an energy equation and a mass equation, taking into account an
ad-hoc heating term which we infer from foot-point UV light curve,
coronal radiative loss, the loss through the transition region which
in this study is scaled to the pressure of the coronal plasma, and
thermal conduction and enthalpy flow between the corona and transition
region.  The synthetic coronal radiation in
SXR and EUV bands are then computed and compared with observations by
GOES and AIA to verify the very few parameters in the method. 

Results of EBTEL calculations for the present datasets are shown in
Figure~8.  Using the 906 UV brightened pixels seen with SDO  we have 
modeled 906 half loops anchored to these pixels. The
lengths of these half loops are estimated from the EUV images, which
range from 16 to 41 Mm.  The heating rates of these loops are assumed to
be proportional to the rise of the UV light curve at the foot-point
pixel with a constant scaling factor which can be adjusted by best
matching synthetic and observed EUV light curves. For this flare, the
inferred heating rate in the 906 half loops ranges from 1$\times 10^{8}$ to 7.5$\times 10^{11}$ ergs
s$^{-1}$ cm$^{-2}$, and the total heating rate is plotted as dashed line in
the top left panel of Figure~8 together with the UV total counts
light curve. By assumption we ignore beam heating scenarios
and assume that (non-radiative) magnetic 
heating primarily occurs in the corona.
The other panels in the figure show synthetic SXR and EUV
light curves in comparison with those observed by GOES 1-8~\AA{} and 6
EUV bandpasses of AIA. Light curves in these passbands reflect
evolution of coronal plasmas that are heated to over 10~MK and then
gradually cool down to 1-2~MK. As shown in Figure~8, the synthetic light curves 
reflect the cooling of the flare, with high temperature emissions
(131 and 94 bands) peaking earlier than low-temperature emissions
(211, 193, and 171 bands). The synthetic and
observed light curves coincide with each other quite well: they
exhibit very similar time profiles and the magnitude of the total counts are
comparable by within a factor of 2. The good comparison indicates that
the method is able to reproduce mean properties of the corona during
the flare.

In these models, the plasma pressure at the coronal base increases from a
pre-flare state between 0.1 and 1 dyne cm$^{-2}$ up to 
pressures of several times $10^{1-2}$ dyne cm$^{-2}$ during the flare.  
Accordingly,
the emission from lines such as \ion{C}{4} increases by similar
factors. The underlying UV continuum near 1600~\AA{}, however, 
forms near pressures closer to
$1-2\times 10^3$ dyne cm$^{-2}$ under quiescent conditions \citep{Vernazza81}, 
over an order of magnitude higher.   Thus, 
we would expect downward propagating 
flare energy to be attenuated before it can reach such high pressures,
associated with the much deeper ``temperature minimum'' region below
the chromosphere.    

Our observations show that the brightest IR 10830, UV 1600, and EUV
304 emission are generated in the lower-atmosphere at the footpoints of
flare loops. Light curves in these wavelengths exhibit an impulsive
rise, followed by a gradual decay. Using the transition region DEMs
computed as above, we find, as a {\em post-facto}
confirmation of our initial assumption, that the UV 1600~\AA{} band is
indeed dominated by the \ion{C}{4} line. The emissivity $\epsilon(T)$ 
(radiated power in erg cm$^{-3}$ s$^{-1}$ divided by N$_e^2$) of the optically thin \ion{C}{4}
multiplet is derived from CHIANTI 7.0 with ionization
equilibrium \citep{Dere97, Landi12}. Using the DEM, the total \ion{C}{4} photon flux is computed in units of photons 
cm$^{-1}$s$^{-1}$sr$^{-1}$ and by convolving with the AIA instrument response
function, the flux is converted to observed count rate in
units of DN~s$^{-1}$. The right panel of Figure~9 shows the synthetic light curve of the \ion{C}{4}
total counts summed up from foot-points of all flare loops, in
comparison with the observed UV 1600~\AA{} total counts. It is seen
that, during the decay phase, the synthetic light curve agrees very well
with the observed light curve in both the evolution timescale and the
magnitude.  However, the impulsive phases are significantly
underestimated by the model calculations, because the equilibrium approximation
is not adequate during the rise (heating) phase of the flare.

The \ion{He}{2} 304~\AA{} line is usually formed in mid transition region
temperature plasma (in equilibrium conditions this would be close to
80,000 K).  To understand the observed \ion{He}{2} emission at the flare
foot-points, we also calculate the contribution function of the
\ion{He}{2} line under conditions of statistical equilibrium, and
convolved it with the transition region DEM and the AIA instrument's 
response function in this band. In Figure~9, the left panel shows the synthetic
\ion{He}{2} total counts light curve, which evolve along with the
observed light curve with comparable amount of emission during the
decay phase.

The successful comparison between the synthetic and observed
\ion{C}{4} light curve during the decay phase confirms that the
pressure-gauge approach reasonably describes the transition region
during this phase.  The similarity of the observed and computed
light curves of \ion{He}{2} 304~\AA{} emission is remarkable, given the
earlier work, particularly that of \citet{Laming92}, cited in the introduction.

\subsection{He I \he{}}

Our observations indicate that the He I \he{} line is formed
somewhere in the stratified chromosphere.  During the flare, this layer
must be significantly heated to enhance the He I line against the
continuum background.  However, the chromosphere is an excellent
thermostat, having enormous sinks of energy associated with latent
heat of ionization of hydrogen and radiation losses.  But if local
heating exceeds a critical limit, then  a thermal 
runaway to coronal temperatures is expected to occur, since once
hydrogen is all ionized, a major sink of energy is lost \citep{Judge10}. 
Dumping of heat into the chromosphere bifurcates the
temperature distribution such that either the temperature is below
$approx$ 8000 K or it moves immediately, with a narrow transition
region, to coronal temperatures.  Since \ion{He}{1} is rapidly ionized
at coronal temperatures, we must look to the underlying chromosphere
for the source of the 10830 photons during the flare.  

Here we consider two pictures: the first is that EUV irradiation of the
chromosphere from above ionized the neutral helium atoms. Then the recombination
 of the electrons and \ion{He}{2} generates populations of the upper levels
($1s2p~^3P^o_{0,1,2}$) of the multiplet \citep{Zirin75}.  The second is
the ``burst model'' advocated by \citet{Laming92}.

The observed enhancements of \he{} range from 1.2
to 2.5 times the pre-flare emission (Figure~6): the median value is
close to 1.3, which we now adopt.  
As mentioned above, most flare footpoints locate in the penumbra. 
Before the flare, the clear view of the sunspot penumbra indicates the shallow absorption in \he{}. 
Therefore, most flux comes from the photosphere in these area. 
So we approximate emission during this stage by the continuum emission. 
The total emission is estimated in physical units from \he{} as follows.   
The estimated continuum formation temperature at \he{} is around 6000 K at disk center as shown by \citet{Maltby86}. 
5600 K is adopted since the flare footpoints locate in penumbra. Assuming that the continuum is generated by black-body radiation, we
can write the excess intensity as 
$$
I_{ex} \approx (1.3-1.0) \times  B_\lambda(5600{\rm K}) \times \Delta \ \ \ \ {\rm erg~cm^{-2} sr^{-1} s^{-1}}, 
$$
Where $\Delta$ is an (unknown) width of the excess emission line
profile in wavelength units.   Let $\Delta = f w$ where $f$ is the
ratio of the radiation emitted by all \he{} transitions divided
by the fraction that we observe, which covers a width $w=0.5$~\AA.  
Values of $1 \le f < 10$ seem
reasonable values, we will use $f=2$ since our bandpass cover almost half of the red component of \he{} spectrum.   We have, using a photospheric radiation temperature near 5600 K = $8\times 10^{5}$ erg~cm$^{-2}$ sr$^{-1}$~\AA$^{-1}$ s$^{-1}$, and $w=0.5$~\AA
$$
I_{ex} \approx (0.3) \, 2\, (0.5)\, 8 \times 10^5  \ \ \ \ {\rm erg~cm^{-2} sr^{-1} s^{-1}}, 
$$
or a total \he{} energy flux out of the optically thin slab of 
$$
F_{ex} = 2\pi I_{ex} \approx 1.5\times 10^6   \ \ \ \ {\rm erg~cm^{-2} s^{-1}}.
$$
Since the energy of one 10830 photon is $1.83\times 10^{-12}$ erg, the
photon flux is then 
$$
N_{ex}  \approx 0.8\times 10^{18}  \ \ \ \ {\rm ph~cm^{-2} s^{-1}}.
$$

\subsubsection{Photoionization-recombination picture}

If the typical \he{} enhancements are produced via the photoionization-recombination mechanism, the required energy flux 
of ionizing EUV photons is simply determined by computing the photoionization rate and the
number of photons emitted in \he{} per ionization \citep[e.g.,][]{Zirin75}. 
Unlike Zirin's work, we need not consider photoexcitation by
the photosphere since the photospheric radiation cannot generate
emission above its own continuum.   
Using the HAOS-DIPER package \citep{Judge07}, we find that
between 32 and 62\% of all
recombinations lead to 10830 emission, depending on whether 
the 504~\AA{} continuum is assumed optically thin or thick
respectively. We adopt a value of 50\% for optical depths close to
unity
where the bulk of the helium material will be photoionized.
EUV photons at
wavelengths below 504~\AA{} are needed for photoionization of  \ion{He}{1}.
We can estimate the photon energy flux at 500~\AA, which is
the minimum required flux to generate the observed enhancement in the
photoionization-recombination picture, as
$$N_{EUV} = N_{ex} /0.5 = 1.6\times 10^{18}    \ \ \ \ {\rm ph~cm^{-2} s^{-1}},$$
for an energy flux of 
$$F_{EUV} \approx  6\times 10^{7}   \ \ \ \ {\rm erg~cm^{-2} s^{-1}}.
$$
When distributed over an area of 906 AIA pixels, we find a total luminosity of the flare at wavelengths
close to and below 504~\AA{} of $1.1\times10^{26}$ erg~s$^{-1}$.  If
distributed across the 9 patches, the average luminosity per patch is
$1.2\times 10^{25}$ erg~s$^{-1}$.   

The solid lines in Figure~10 shows the estimated photon energy flux from the observed IR
enhancement in each of the 8 patches. P5 is not plotted since its emission is blended with 
the absorption from the nearby filament. With the transition region DEM computed for every
pixel, we estimate the total optically-thin radiation energy by
transition region plasmas of up to 2~MK, as plotted in dotted lines in
Figure~10.  It is seen that, in most of these patches, the estimated
radiation energy is comparable with the required photon energy to
generate the enhanced IR emission during the decay phase for as long
as observed. We conclude that 
photoionization by EUV photons in the overlying transition region is a
viable mechanism for the prolonged He I emission at the flare
foot-points.

An independent EUV photon energy flux produced in this
flare can be estimated from observations by the 
Extreme Ultraviolet Variability Experiment \citep[EVE;][]{Woods12}
 on SDO. Two EVE spectrographs measure the whole-Sun 
solar extreme ultraviolet
(EUV) radiation spectrum from 10-1050~\AA{} with a resolution of
approximately 1~\AA{} and a cadence of 10 seconds. During this flare
the ``A'' spectrograph on EVE acquired data in the range 
 60-380~\AA{}, no data above 380~\AA{} were obtained with the ``B''
 spectrograph.  The net EUV flux from the flare, obtained by subtraction
of pre-flare emission, 
is shown in Figure~11. The solid line is the total flux observed by EVE from
60-380~\AA{} while the dotted line is the calculated total optically-thin
radiation by plasmas of up to 2~MK. The observed and measured fluxes
are within $\pm$ a factor of three. Given the crude nature of the 
calculations and 
complex geometry of the flaring plasmas at various wavelengths, 
these measurements provide reassurance that the photon fluxes in
the model are broadly compatible with the data. 

\citet{Laming92} estimate $2-8 \times 10^{16}$ photons
cm$^{-2}$~s$^{-1}$~sr$^{-1}$ for the EUV ionizing intensity in the \ion{He}{1}
and \ion{He}{2} continua. The number directed towards the chromosphere is
$2\pi$ larger, or $\sim 3\times10^{17}$  photons cm$^{-2}$~s$^{-1}$.
This is a factor of 7 smaller than our estimate. This difference may
be real, reflecting different flares, or it may reflect differences in
the resolution of instruments used. Our data has a far higher angular
resolution, necessarily leading to higher intensities as smaller flare
kernels are better resolved.

\subsubsection{Burst picture}

The essence of this picture is that a burst of heating is assumed to
occur in cool (chromospheric) plasma on time scales small compared
with the time taken to ionize a given ion.  During the burst the
electron
temperature is raised above the quiescent state, the hotter electrons
lead to line emission before the ions involved become ionized.
Heuristic arguments for
this model have been given for EUV/UV helium lines by \citet{Laming92}.
 It has been placed on firmer theoretical grounds by
\citet{Judge05}.  In the work of Laming and Feldman, burst temperatures
near $1.8 \times10^5$ K were needed to reproduce spectra of
\ion{He}{2} which are consistent with spectra of \ion{He}{1}.

This picture seems unlikely to be able to explain the \he{}
line's behavior reported here for the following reasons.  The
\ion{He}{1} ionization times for pre-flare electron densities of
around $5\times 10^{10}$ cm$^{-3}$ \citep[the value in the pre-flare upper
chromosphere from][]{Vernazza81} are about 100, 2
and 0.05 s for burst electron temperatures of $2.3\times 10^4$,
$4.6\times 10^4$, and $9.3\times 10^4$ K ($\equiv$ 2, 4 and 8 eV),
respectively.  These values are simply 
scaled from Laming \&{} Feldman's Figure~1.  Excitation
of the $1s2p~^3P^o$ levels by electron collisions naturally requires time
scales with a similar temperature 
dependence because these levels lie at 21
eV, close to the continuum which is at 24.5 eV.  Thus, populating
these levels also leads to significant ionization. Further, 
a mere 1-2 eV
is sufficient to ionize 50\% of \ion{He}{1} when photoionization from
the $n=2$ levels is considered (Laming \&{} Feldman table 4).  
There are two difficulties:  there is ``little room in temperature space'' where electron
collisions from below can provide significant populations of the 
$1s2p~^3P^o$ levels- if they are excited, the \ion{He}{1} is also readily 
ionized; for electron temperatures equivalent to 4 eV or more, needed
to provide significant populations of these levels, 
the ionization times are so short so that one must invoke very many
bursts to maintain the observed light curves, even during the
``impulsive'' rise phases (many tens of seconds for 10830~\AA, see Figure~7). 

Thus, recombination appears the only way to significantly populate the
$1s2p~^3P^o$ levels.  If this occurs via EUV photons, helium is
selectively enhanced by the process outlined above. If however
electron collisions are responsible, then one would expect very bright
UV and EUV emission from trace species such as C, N, O, Si\ldots{}
from the upper chromosphere.  Simultaneous 
observations of UV emission lines of
various ions, for example from the new IRIS spacecraft, would be
needed to see if pictures involving electron impact excitation of
\ion{He}{1} from the $1s^2 ~^1S$ level can be refuted or must be
considered further.

\section{DISCUSSIONS AND CONCLUSIONS}

With the NST's high spatial resolution imaging observations as well as
simultaneous space data from AIA on board SDO, we have given a
detailed analysis of a C class flare observed continuously by all
instruments at the cadence of 10-24~s (30~s for the TiO NST data). 
Specifically, we have tried to identify the
dominant mechanism that produces the observed \ion{He}{1} 10830~\AA\
emissions.

We adopt the picture that the flare energy released in the corona
propagates downwards by heat conduction and radiation into the
underlying chromosphere.  Particle beams seem insignificant on the basis
that HXR fluxes are very small for this flare. 
Furthermore, our analysis focus on the decay phase of the flare, when the 
thermal model is sufficient for fitting the RHESSI spectra \citep{Liuwjphd14}.
We compute the conductive downflow by
determining heating rates inferred from 906 UV brightened pixels and
the zero-dimensional EBTEL model to compute mean plasma properties in
the corona and transition region of flare loops anchored within these
pixels. Remarkably, the method can reproduce not only the \ion{C}{4}
 emission generated in the transition region at the flare
foot-points, but also \ion{He}{2} 304~\AA{} emission, at least during
the decay phases, to within a factor of two or so.  This is remarkable
agreement given the need to appeal to different mechanisms for bigger
flares in previous work \citep{Laming92}.

In terms of the \ion{He}{1} \he{} multiplet, we argue that the
downward propagation of energy via EUV photons is an important
mechanism for excitation during the flare.  Thus photo-ionization
followed by recombination \citep[PR;][]{Zirin75} appears to be a prominent
component exciting the \he{} multiplet during this event.  Several
pieces of evidence support this picture.  Firstly, the morphology
of the footpoint \he{} emission is qualitatively similar to a mixture
of EUV channels observed with AIA.  Secondly, the light curve of \he{}
emission is, during the rise phase, similar to the (E)UV transition
region light curves (304~\AA{} and \ion{C}{4} which dominates the 1600~\AA{} channel); 
during the decaying phases, it is perhaps more similar
to the coronal SXR and EUV light curves.  Thirdly, photon budgets are in
reasonable agreement, as measured against both EVE data as well AIA
data.  Fourthly, the EUV radiation computed from the EBTEL models is
also compatible with the photon budget for exciting \he{} via PR,
during the decay phase.

In \citet{Andretta08}, the calculated helium spectrum is strongly affected by the spectral distribution and overall level of EUV coronal back-illumination, which is consistent with our result. They also found that \ion{He}{1} \he{} is not sensitive to chromospheric helium abundance. In our calculation the helium abundance merely needs to be large enough so that it dominates
most of the absorption of some EUV radiation, these photons being converted to longer wavelengths such
as the \he{} line we computed down to the layer where EUV no longer penetrate. Our calculations
are therefore insensitive to modest changes in the  abundance of helium.

But there is perhaps one problem with the PR mechanism: if this were
the {\em only} source of photons in \he{}, then there should be
a close correspondence between the spatial distribution of the EUV
photoionizing radiation and the underlying \he{} flare
emission.   However, this is not always the case. 
Scatter plots of peak intensities of the light curves in AIA 1600~\AA, AIA 304~\AA\ and \ion{He}{1} \he{} show little correlation. 
One cause of the enormous scatter might be the fact that we
observe only a 0.5~\AA{} wide part of the \he{} line, shifted by 0.25~\AA{} 
blueward of the line center. Therefore the morphology and
light curves might contain velocity-dependent information that is not
present in the broad band (E)UV channels.  

Nevertheless, we also examined a ``burst'' picture which depends on
non-equilibrium ionization to explain the large populations of the
$n=2$ triplet levels of helium.   We concluded that this seems unlikely
for the \ion{He}{1} \he{} multiplet on several grounds, at least for
the decay phases.   Occam's Razor would compel us to accept the PR
mechanism as perhaps the ``simplest answer compatible with the data''.

The assumed steady-state and equilibrium conditions are not adequate
for calculations during the impulsive phases with durations of 10-100
seconds or so.  In this period, the \he{} have identical rapid
enhancement with the (E)UV lines.   Perhaps a ``burst'' picture would
be successful during this phase.  Further observations
including more UV lines, such as can be observed with IRIS, would be
desirable, since to leading order 
the lines would respond to bursts but not the PR mechanism.

\acknowledgements {We thank BBSO observing staff and instrument team for their support. We thank Dr. Haisheng Ji for providing the narrow-band \he{} Lyot Filter. We are grateful to Giuliana de Toma for a careful reading of the paper. SDO is a project of NASA. Z. Z. and W. C. acknowledge the support of the US NSF (AGS-0847126, AGS-1146896 and AGS-1250818) and NASA (NNX13AG14G). J. Q. acknowledges the support by NSF (ATM-0748428) and NASA (NNX14AC06G). P. J. acknowledges the support by Montana State University Physics department, and the ASP and HAO programs at NCAR. NCAR is sponsored by the National Science Foundation. We thank the referee for carefully reading our paper and for providing valuable suggestions which substantially helped improving the quality of the paper.}

\bibliography{ref}

\newpage
\begin{figure}
\epsscale{0.9} \plotone{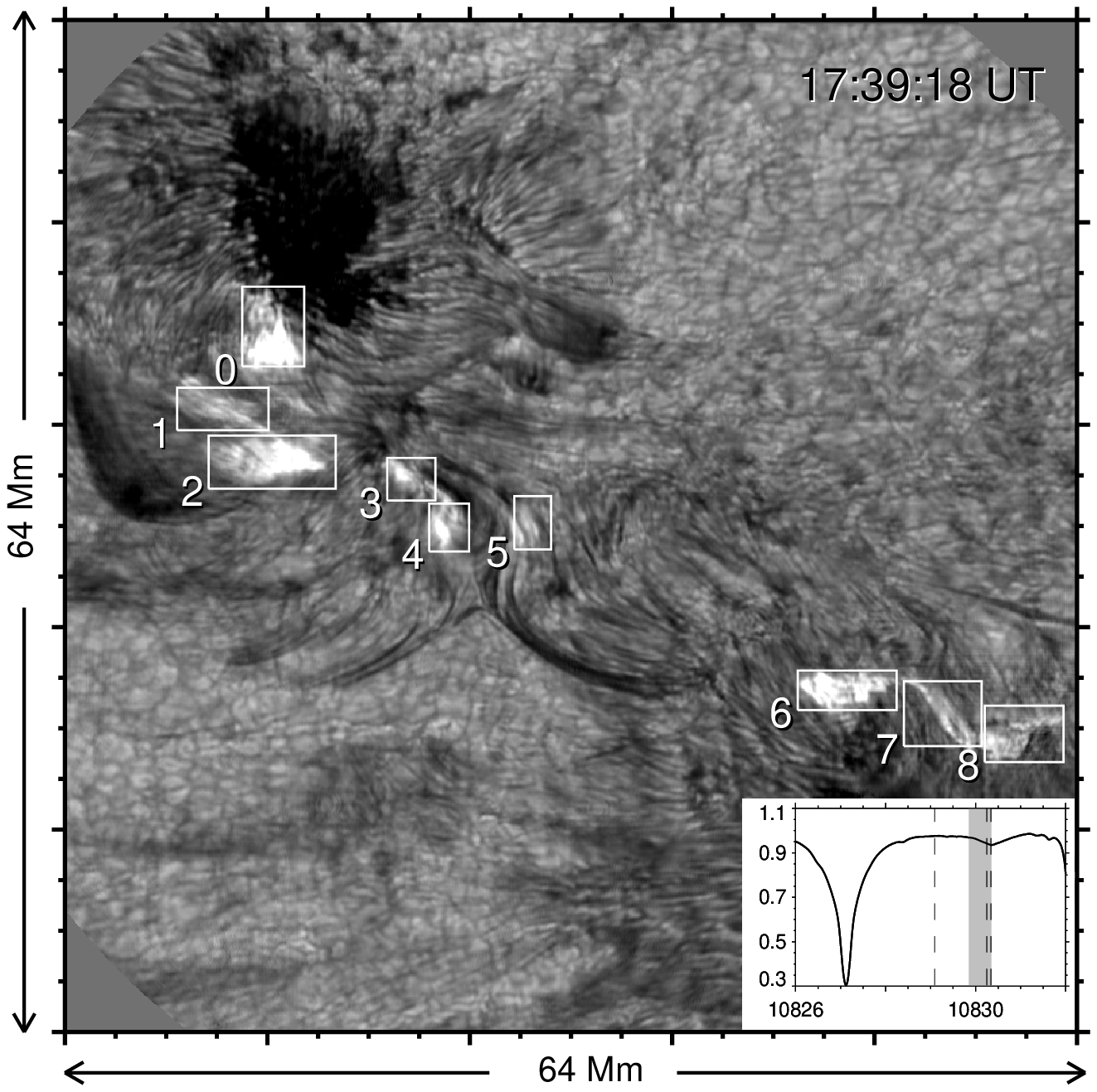}
\caption{Snapshot of the \ion{He}{1} \he{} observation. The bandpass of the narrow-band filter is illustrated by a vertical strip in the solar spectrum on the lower right corner, and three vertical dashed lines depict the line centers of the \ion{He}{1} 10830 \AA\ triplets. The field of view is 64 Mm $\times$ 64 Mm. The 9 small boxes with digits encompass the footpoints of the flare during its second peak.  }
\end{figure}

\begin{figure}
\epsscale{0.9} \plotone{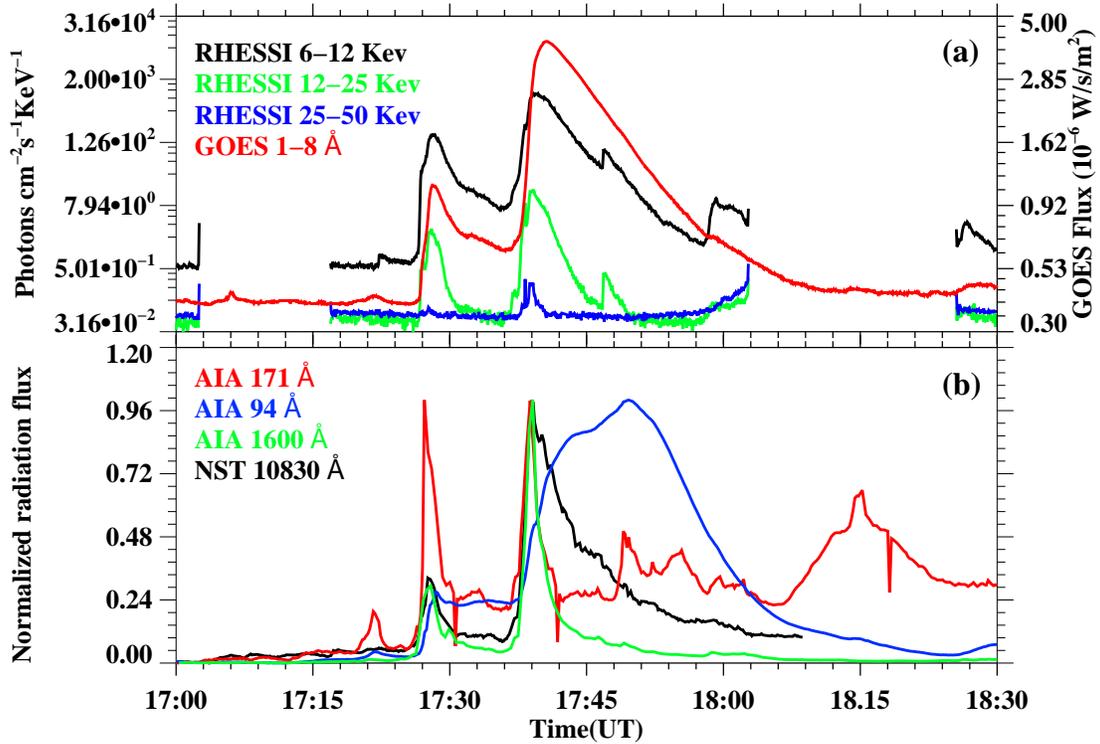}
\caption{Light curves of the 2012 June 17 C3.9 flare. Both soft X-ray 1-8~\AA{} from GOES and some X-ray flux from RHESSI are shown in panel a. Time profiles in UV 1600~\AA{} and EUV 171~\AA{} and 94~\AA{} by AIA/SDO are plotted in panel b. Also plotted is the light curve of 10830~\AA.}
\end{figure}

\begin{figure}
\epsscale{0.9} \plotone{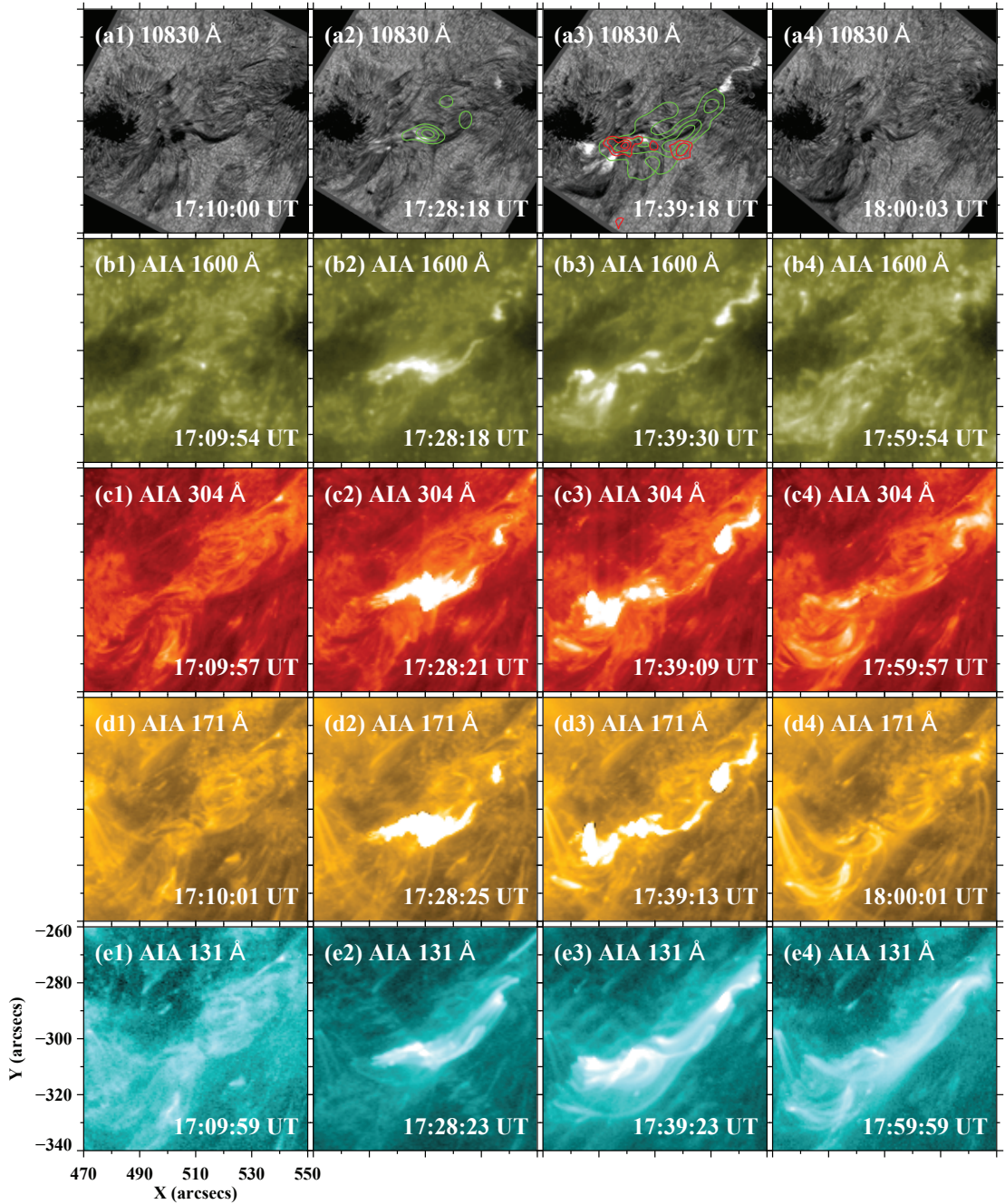}
\caption{Snapshots of the flare in a few bands at a few moments with the same field of view. Rows from top to bottom: \ion{He}{1} 10830~\AA, UV 1600~\AA, EUV 304~\AA, 171~\AA{} and 131~\AA. Left panels (a1-e1): snapshots before the filament eruption. Middle panels (a2-e2): snapshots during the first peak of the flare. Panels (a3-e3): snapshots during the second peak of the flare. Right panels (a4-e4): snapshots after the flare. The contours on the \he{} filtergrams are from RHESSI images with channels of: 6-15~keV (green) and 25-50~keV (red). The RHESSI images are made from 17:28:10 UT with 10 seconds integration and from 17:38:30 UT with 1 minute integration, respectively. The contour levels are 90\%, 75\% and 60\% of the peak intensity of each image. }
\end{figure}

\begin{figure}
\epsscale{0.9} \plotone{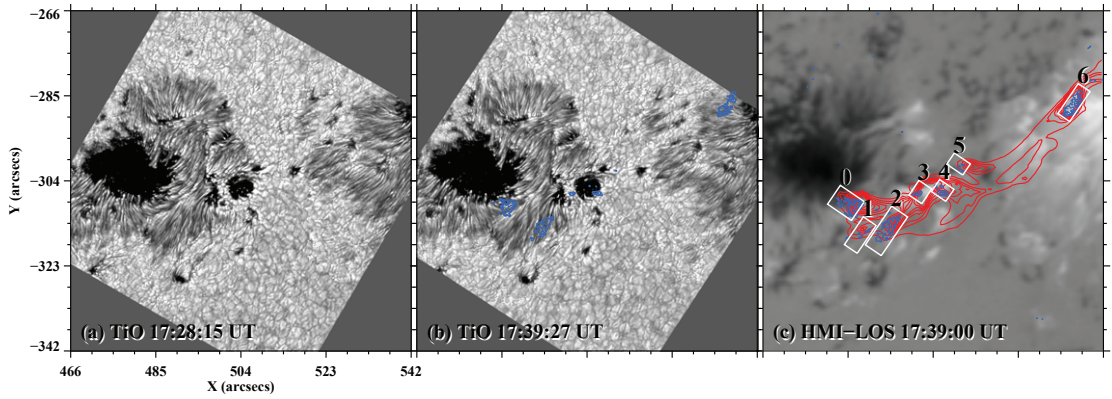}
\caption{Panel (a) and (b): TiO images from NST/BBSO during the two impulsive phase. Blue contours on panel (b) are regions of emission measured in \ion{He}{1} 10830~\AA. Panel (c): HMI line-of-sight (LOS) magnetogram overlaid with foot-point contour in \he{} (Blue) and loop contour in 131~\AA{} (Red). The boxes encompass the patches indicated in Figure~1. Three panels have the same field of view of 76$\arcsec$ $\times$ 76$\arcsec$. }
\end{figure}

\begin{figure}
\epsscale{0.9} \plotone{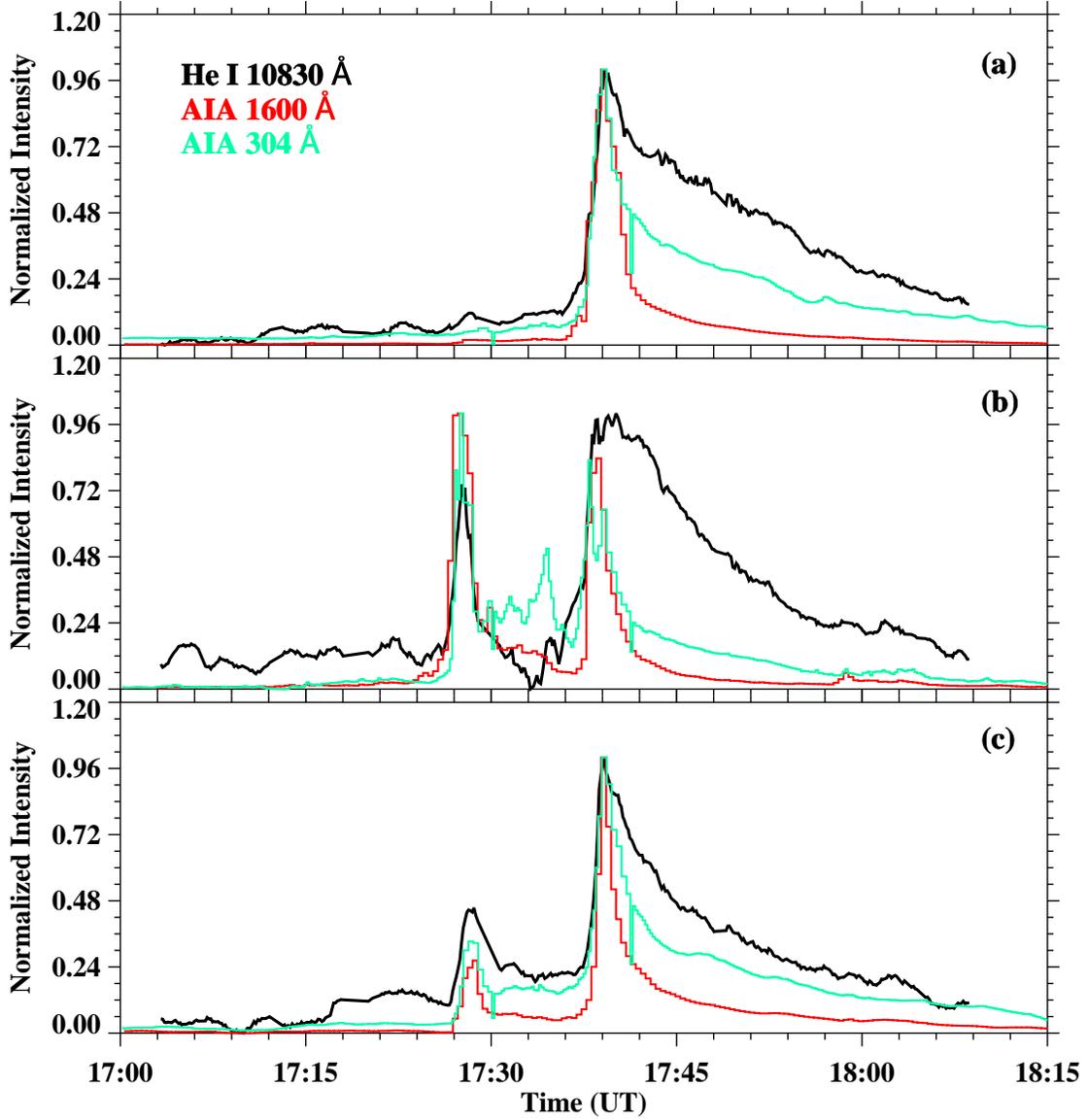}
\caption{Light curves in \ion{He}{1} 10830~\AA, AIA 1600~\AA{} and 304~\AA. Panel a to c represent for patches 0, 4 and 6 noted in Figure~1, respectively. The minimum value is subtracted from each light curve, which then normalized to its maximum.}
\end{figure}

\begin{figure}
\epsscale{0.9} \plotone{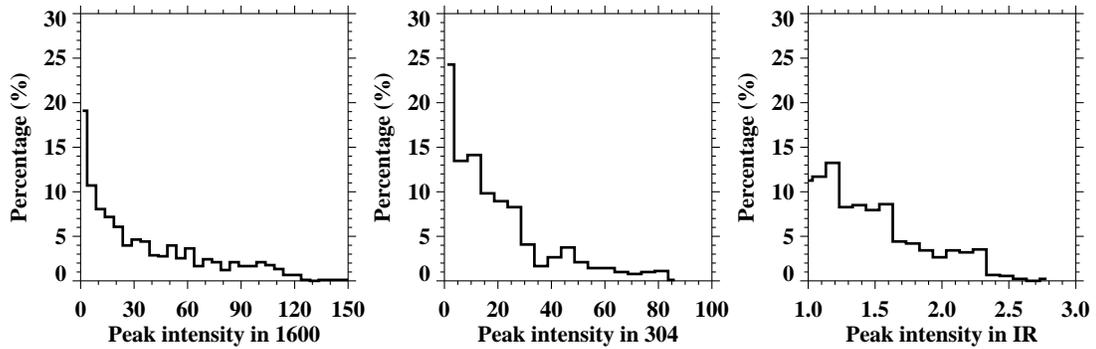}
\caption{Histogram of peak intensities for all the footpoint pixels: x axis shows peak intensities as how many times the pre-flare intensity; y axis is in percentage. For the \ion{He}{1} \he{}, the pre-flare intensity is approximated by the quiescent continuum.}
\end{figure}

\begin{figure}
\epsscale{0.9} \plotone{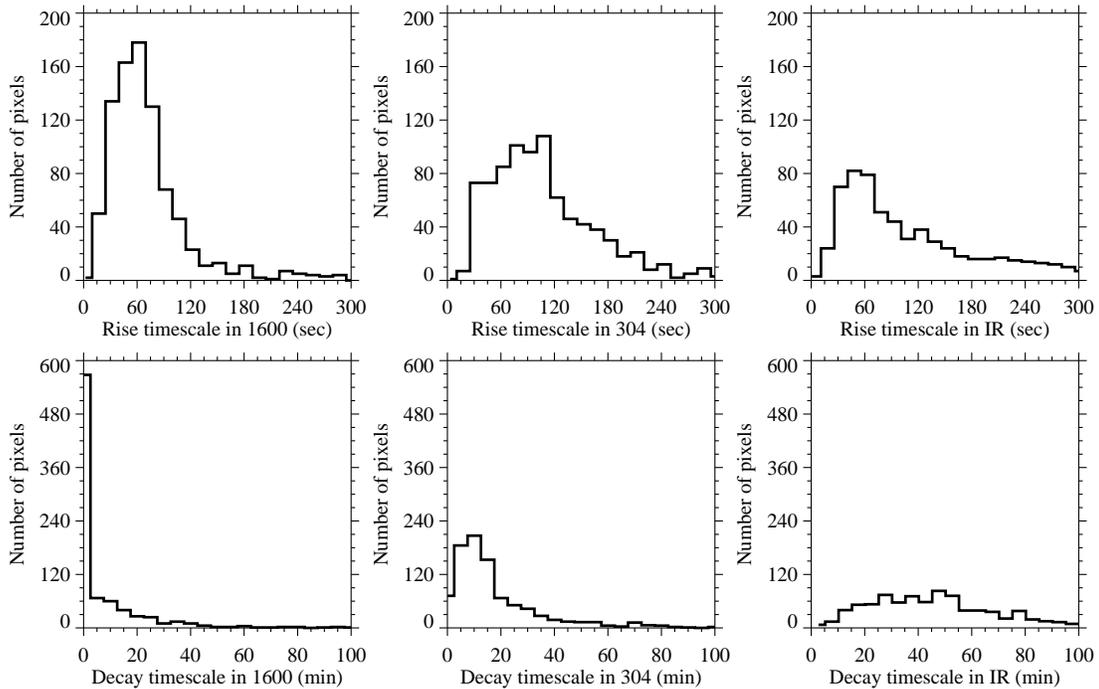}
\caption{Histogram of 10830~\AA, 1600~\AA{} and 304~\AA{} rise and decay timescales. Upper panels: rise timescales. Bottom panels: decay timescales.  }
\end{figure}

\begin{figure}
\epsscale{0.9} \plotone{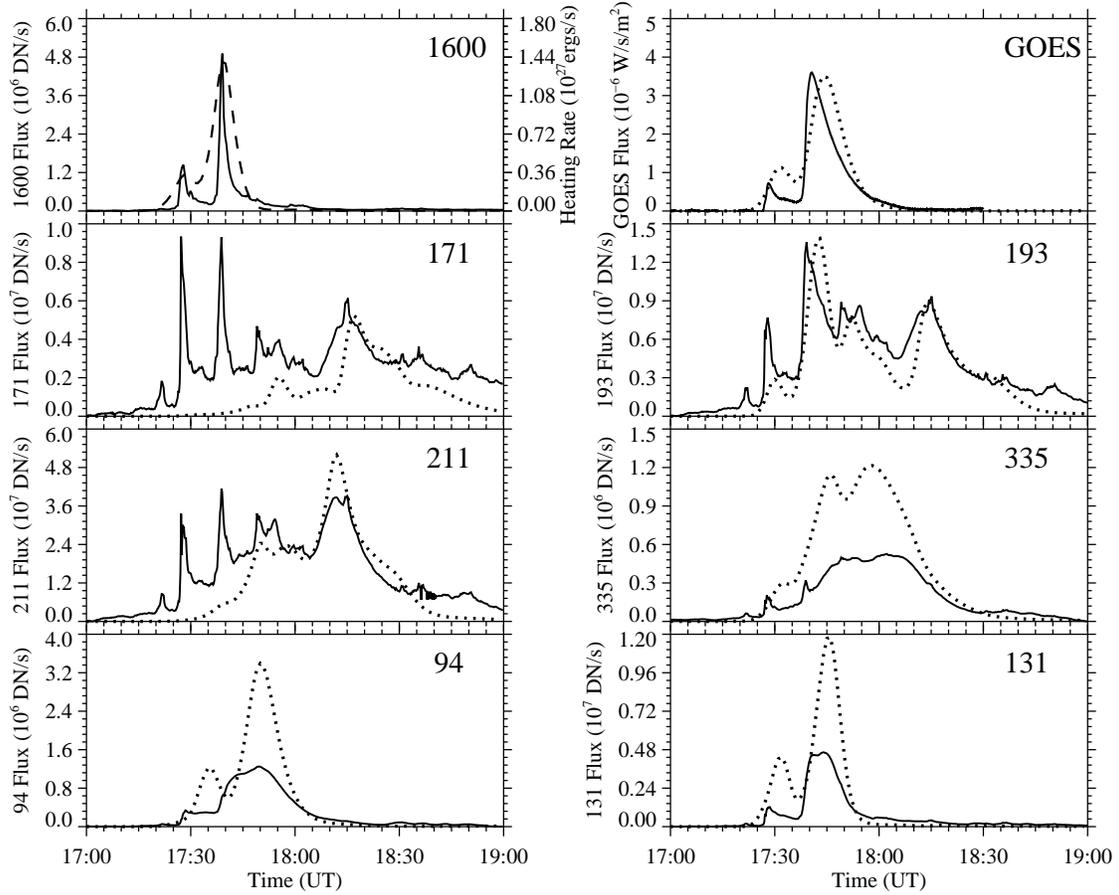}
\caption{Comparison of observed and synthetic flux calculated through EBTEL model. The dotted and solid lines are the synthetic and observed light curves, respectively. The heating rate is given as the dashed light curve in the top left panel. The corresponding passbands are noted by the number on the right corner of each panel.}
\end{figure}

\begin{figure}

\epsscale{0.9} \plotone{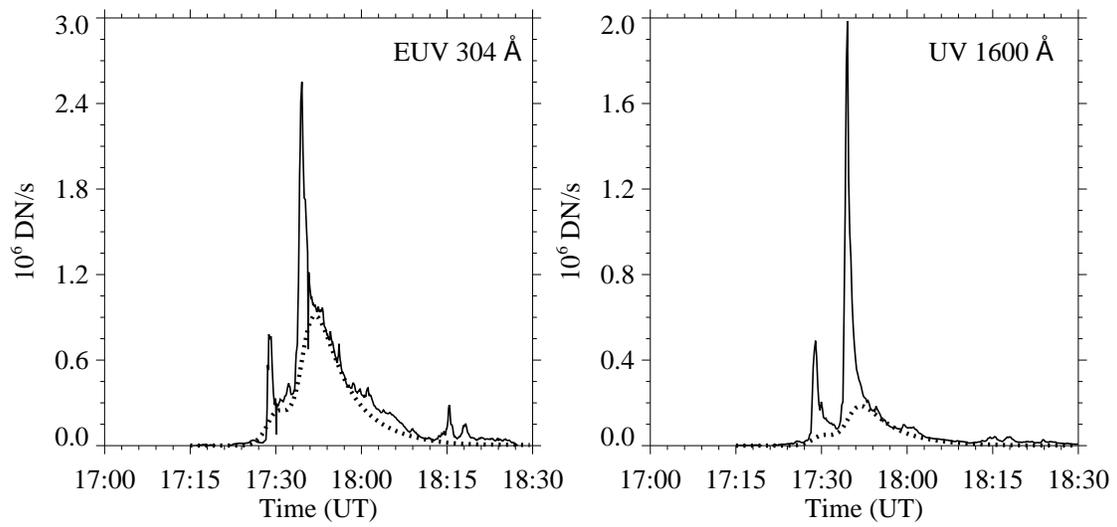}
\caption{Dotted lines: computed light curves for 304 and 1600~\AA{} with pressure gauge assumption. Solid lines: observed data.}
\end{figure}

\begin{figure}

\epsscale{0.9} \plotone{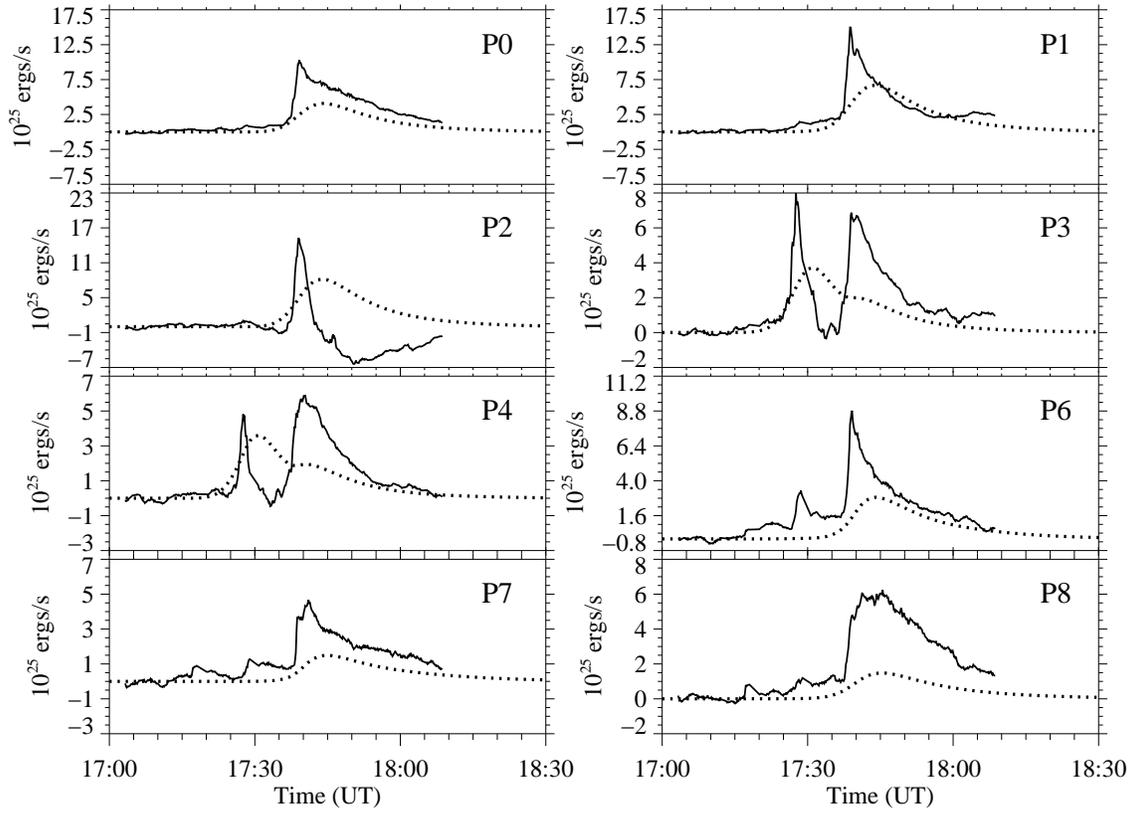}
\caption{Total estimated optically-thin energy flux from the transition region (dotted line) compare with estimation from \he{} enhancement (solid line). The number on the upper right conner of each panel denotes the patches showed in Figure~1. The patch P5 is not plotted here because the emissions are blended with the filament absorption. }
\end{figure}

\begin{figure}

\epsscale{0.9} \plotone{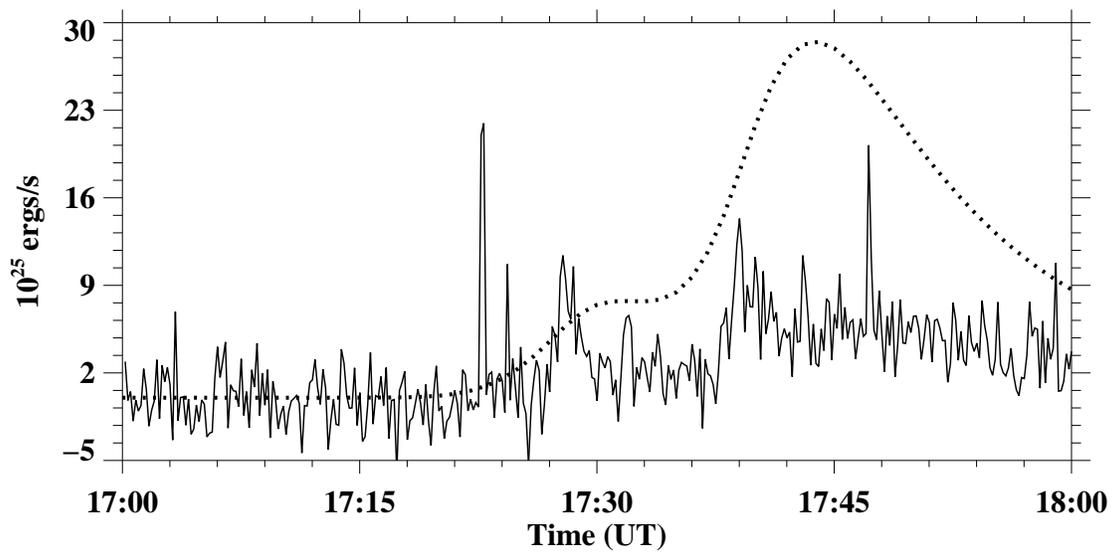}
\caption{Solid line: EVE observed EUV flux summed up through the spectrum range from 60-380~\AA{} with pre-flare emission subtracted. 
Dotted line: model estimated total optically-thin radiation flux with pressure-gauge approximation.}
\end{figure}

\end{document}